\documentclass[12pt,aps,preprint,epsfig,shoswpacs]{revtex4}
\usepackage{graphicx}
\usepackage{epsfig}
\begin{document}
\title{ Does changing the pulling direction give better 
insight into biomolecules? }
\author{Sanjay Kumar and Debaprasad Giri$^\dagger$}
\affiliation{Department of Physics, Banaras Hindu University,
Varanasi 221 005, India \\
$^\dagger$Physics Section, MMV, Banaras Hindu University,
Varanasi 221 005, India}

\begin{abstract}
Single molecule manipulation techniques reveal that the mechanical
resistance of a protein depends on the direction of the applied 
force.  Using a lattice model of polymers, we show that changing 
the pulling direction leads to different phase diagrams. 
The simple model proposed here indicates that in one case the system 
undergoes a transition akin to the unzipping of a $\beta$ sheet, while 
in the other case the transition is of a shearing (slippage) nature.
Our results are qualitatively  similar to experimental results. 
This demonstrates the importance of varying the pulling direction since this
may yield enhanced insights into the molecular interactions responsible 
for the stability of biomolecules.
\end{abstract}
\maketitle

The last decade has witnessed an intense activity in experiments 
involving the manipulation of  single biomolecules. This interest has
been fueled on the one hand by the desire to understand the fundamental 
mechanisms at play in biological systems, and on the other 
hand by the development of revolutionary single-molecule force 
spectroscopy experiments \cite{rief,busta,busta1,Oberhauser,
Marszalek}.  These experiments provide
unexpected insights into the strength of the forces driving biological 
processes and help to determine various biological interactions as well as
the mechanical stability of biological structures. 
In some cases the experimental setup also allows one to 
locate precisely the positions of the forces occurring within 
the biomolecule \cite{mueller}. 

The theoretical studies (numerical and analytical) which
followed the experimental efforts
have mostly been confined to modeling the molecule within the 
context of statistical mechanics. The models use
various kinds of simplified interactions and compare
the resulting theoretical predictions against the experimental 
findings. For example, the most widely used models are 
the freely jointed chain (FJC) and worm like 
chain (WLC) models \cite{fixman,doi}, which describe the 
force-extension curves in the intermediate and high-force regimes. 
However, both these models ignore
crucial excluded volume effects \cite{degennes}, and are 
thus only well suited to modeling the stretching of proteins  in 
a good solvent. Note that solvents relevant in a biological context 
are usually poor. Therefore, these models and numerical studies 
(Monte Carlo simulations) are unable to access the low temperature regime 
relevant in a biological context. Consequently the study of the 
emergence of intermediate states stabilized by a force at low 
temperature is beyond the scope of these models. 

\begin{figure}[t]
\includegraphics[width=2.0in]{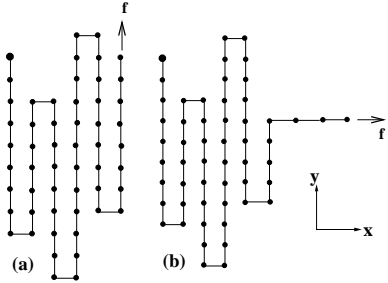} \\
\caption{\label{fig:model} Schematic illustrations of PDSAWs on 
the square lattice. One end is fixed 
and the other end is subjected to a pulling force (a) perpendicular to
the preferred direction ($y$-direction); (b) along 
the preferred direction ($x$-direction).}
\end{figure}

Efforts have recently shifted to the experimental study of molecular
conformations of biopolymer by changing the pulling direction 
\cite{yang,Dietz,brock}. For example, in a recent experiment, Dietz and 
Rief \cite{Dietz} showed that by changing the
pulling direction (mechanical triangulation) one obtains  distinct
force-extension curves from which angstrom-precise
structural information can be obtained about single proteins
in a solution. Neither the FJC or WLC models nor 
the self-avoiding walk (SAW) model (which does include excluded
volume effects \cite{degennes,vander,kumar}) shows any of 
the effects related to a change in the
pulling direction contrary to the observations of recent experiments
\cite{Dietz,brock}. This is  because the 
shape of the chain conformations and the interactions  seen in 
these models are isotropic in nature. Notably 
all the proteins studied so far are highly anisotropic both in shape and 
interactions.

In this letter, we study the force-extension curves of flexible
and semi-flexible polymers by changing the direction of the pulling force
using an inherently anisotropic lattice walk model. 
We show that a change in the pulling direction \cite{Dietz,yang,brock} 
gives rise to many new intermediate states and that the force-temperature 
phase diagrams are significantly different. In order to model  
the anisotropy of the biomolecules, we use a lattice model
of partially directed self-avoiding walks (PDSAWs) in which
steps with negative projection along the $x$-axis are forbidden \cite{vander}.
At low temperature and high stiffness, the model mimics the structure of 
$\beta$ sheets \cite{kumar} as seen in molecules like titin \cite{rief,busta}.
The major advantage of the model is that it can be solved exactly in the 
thermodynamic limit.  In all single molecule experiments, a chain of finite 
size has been used and hence in principle no ``true phase transition" can  be 
observed \cite{giri}. In order to study finite-size effects, it is 
essential to study first chains of finite length and then their thermodynamic 
limit.
The PDSAW model can be solved exactly in the 
canonical ensemble for finite chain lengths $N$, and using finite-size data, 
the thermodynamic limit of the model can be extrapolated and compared with 
values obtained from exact solutions \cite{rajesh, rosa}.

The model of PDSAWs on a two dimensional square lattice 
is shown in Fig. 1. 
The stiffness  of the chain is modeled 
by associating a positive energy ($\Delta$) with each turn or bend 
of the walk \cite{kumar}. For a semi-flexible
polymer chain the extended state may be favored by increasing the 
stiffness. The stretching energy $E_s$ arising due to the
applied force $f$ is taken as
$E_s = -f\cdot \alpha$, 
where $\alpha$ is the $x$-component (or  $y$-component) of the end-to-end
distance $(|x_1-x_N|)$ (or $(|y_1-y_N|)$). This distance has 
been used as the mechanical reaction coordinate that monitors 
the response of the force \cite{busta1} and it gives important information 
about the conformation of the biomolecules.

\begin{figure}[t]
\includegraphics[width=3.2in]{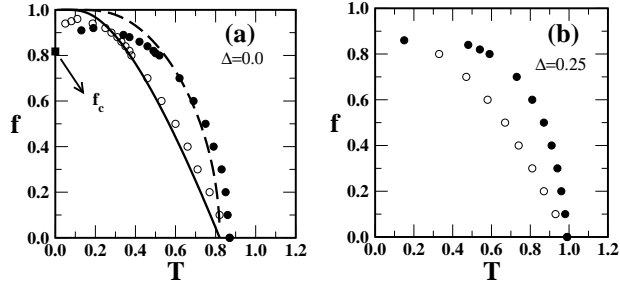} \\
\caption{\label{fig:phase} The globule-coil phase boundary in the
force-temperature plane: (a) for flexible polymer chains; 
(b) for semi-flexible polymer chains.
The phase diagram corresponding to a force applied along the
$y$-direction (filled circle: finite $N$ and dashed line: exact 
phase boundary \cite{rajesh}) is distinctly different from that of a force 
along the $x$-direction (open circle: finite $N$ and solid line:
exact phase boundary \cite{rosa}).} 
\end{figure}

The complete partition function of the system under consideration
can be written as
$  Z_N(N_b, \sigma, | \alpha |)  = \sum_{(N_b, \sigma, 
    |\alpha|)} C_N (N_b, \sigma, |\alpha |) s^{N_b} 
  a^{\sigma} p^{| \alpha |}$, where
$C_N (N_b, \sigma, | \alpha |)$ is the total number of PDSAWs
of $N$ steps having $N_b$ turns (bends) and $\sigma$ nearest neighbor
pairs;  $p$ is the Boltzmann weight for the
force defined as $\exp[\beta (f.\hat{\alpha})]$,
where $\hat{\alpha}$ is a unit vector along the {\em x}-axis
(or {\em y}-axis); $a =\exp[-\beta \epsilon]$ and $s =\exp[-\beta 
\Delta]$ are the Boltzmann weights associated with
nearest neigbour interactions between non-bonded monomers 
and bending energy, respectively.
We use the exact enumeration technique to find $C_N$ for chains
of length up to $N =30$ and analyze the partition functions.
Scaling corrections can be taken into account by suitable 
extrapolation schemes enabling us to obtain accurate estimates in the 
thermodynamic (infinite length) limit \cite{vander}. 
The reduced free energy per monomer is found from the relation
$G = \lim_{N \to \infty} (1/N) \log Z_N((N_b, \sigma, | \alpha |))$.
The limit $N \to \infty$ is achieved by using the ratio method
\cite{vander} for extrapolation. The transition point for
flexible chains ($\Delta= 0$) at zero force ($p=1$),
{i.e.} a coil-globule transition, can be obtained either from a plot of
$G$ versus $a$, or from the peak value of
$\frac{\partial^2 G}{\partial (\ln a) ^2} $.
We find $a=3.336$ at $p=1$. This is shown (in terms of
temperature, $T=0.83$) in the force-temperature (f-T) phase diagram.
The force and temperature are obtained from the expressions for the 
Boltzmann weights $f=\log(p)/\log(a)$ and $T=1/\log(a)$,
respectively, by setting $\epsilon=-1$. This value is in excellent 
agreement with the exact value ($T=0.8205$) \cite{rosa}.
Moreover, this value is also quite close
(within error bars of $\pm0.02$) to the one  obtained from
the fluctuations in non-bonded nearest neighbors (which can also be
calculated exactly for finite $N=30$).
At zero force, the system attains the globule (folded or $\beta$ sheet) 
state as shown in Fig.~1 below $T <T_c$. 
The qualitative behavior of the phase diagram (shown in Fig. 2) 
is similar to the one reported in \cite{kumar} for SAWs. It should
be noted that for finite $N$, as well as in the thermodynamic limit, the 
phase diagrams obtained when the force is applied 
along the $y$-direction are distinctly different from the 
corresponding phase diagrams with the force applied along the $x$-direction.  
Reduced temperature and force may be expressed in real units 
by using the following expressions: $T_{exp}=\epsilon_{exp} T/k_B$
and  $f_{exp}=\epsilon_{exp} f$. Here, $k_B$ is the Boltzmann constant and 
the subscript ``{\it exp}" corresponds to values in real units. 
For example, if one chooses $\epsilon_{exp} = 1$ kcal/mol, then the 
equivalent force will be of the order of $70$ pN nm. 

\begin{figure}[b]
\vspace {.2in}
\centerline{\includegraphics[width=3.5in]{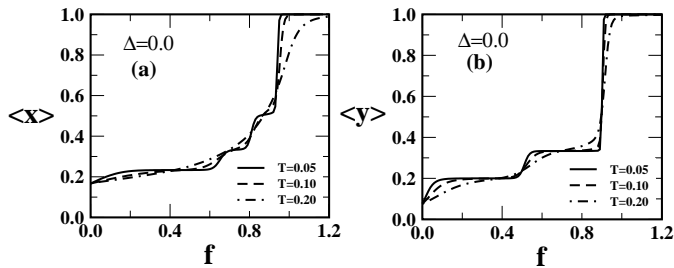}}
\caption{\label{fig:force_ext1} (a) The average scaled extension
of a flexible polymer chain as a function of the pulling force 
$f$ at different temperatures : (a) along the $x$-direction; 
(b) along the $y$-direction.}
\end{figure}

Remarkably, for {\em finite} $N$ the force-temperature phase
diagram shows re-entrance for flexible chains, but re-entrance is absent for
semi-flexible chains. However, in the thermodynamic limit, there
is no re-entrance \cite{rosa}  for flexible  chains.
The presence of re-entrance (at finite $N$) may be explained
by using a phenomenological argument near 
$T=0$. The dominant contribution to the free energy, 
\begin{equation}
G (\equiv -f N) =  N \epsilon - 2 \sqrt{N} \epsilon - N T S_c
\end {equation}
is from the first term.
The second term is due to surface corrections which  
vanishes in the thermodynamic limit, but plays a very important role 
for finite $N$.  The last term is a contribution due to the entropy 
associated with the globule, where $S_c$ is the entropy per monomer. 
It may be noted that for PDSAWs at $T=0$, there are only two 
conformations (Hamiltonian walks) which are the most 
compact configurations and hence one does not see any re-entrance
in the thermodynamic limit \cite{rosa}. For $T >0$, there is a 
finite entropy associated with the deformed globule, which along 
with the surface correction term gives rise to re-entrance in the finite 
chain. The critical force, $f_c$, for $N=30$ found from Eq.~(1) is 
equal to $0.8174$ at $T=0$. This value is indicated
by a black square on the $y$-axis of Fig. 2. This is less than 
$1$ \cite{rosa} as obtained in the thermodynamic limit from Eq.~(1)

In Figs.~3 and 4, we plot the the average scaled extension  
for flexible and semi-flexible polymer chains, respectively, by
using the expression:
$\langle \alpha \rangle / N = (1/N) \sum \alpha C(N_b,\sigma,\alpha) a^\sigma
 p^\alpha/\sum C(N_b,\sigma,\alpha) a^\sigma p^\alpha$.
The extension-force curves show multi-step transitions at low 
temperature corresponding to intermediate states. In the constant 
force ensemble, there is an additional contribution  
to the free energy proportional to the product of the force and the extension 
(along the direction of the force). This contribution stabilizes the 
intermediate states of the globule and hence the observed multi-step behavior. 
Multi-step transitions have also been observed
in recent experiments \cite{haupt} where the globule deforms into 
an ellipse and then into a cylinder. At a critical extension, the 
polymer undergoes a sharp first order transition into a ``ball 
string" conformation \cite{haupt}. This shows that finite 
size effects are crucial in all single molecule experiments 
\cite{lemak}. When the temperature is
increased the multi-step character of the 
extension-force curve is washed out due to increased contributions 
from the entropy \cite{maren}. This effect can be seen in Fig.~3. 

\begin{figure}[b]
\vspace {.2in}
\includegraphics[width=3in]{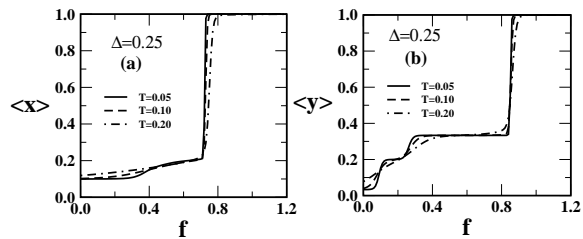}
\caption{\label{fig:force_ext2} Same as Fig. 3, but for the 
semi-flexible chain. }
\end{figure}
In contrast to the FJC, WLC or SAWs, in PDSAWs the walk 
is directed along the $x$-direction and it is inherently anisotropic
so that the perpendicular 
and parallel components scale differently, namely 
as $\sqrt{N}$ and $N$, respectively \cite{vander}. 
Hence the phase boundaries for these two cases remain 
distinct even in the thermodynamic limit, as shown in 
Fig.~2. Here, one can also see that in order to unfold the chain 
at a given temperature, one needs a stronger force along the $y$-axis 
than along the $x$-axis. 

In the case of a semi-flexible chain,  
the response of the force is more pronounced  and the emergence of
intermediate states by changing the pulling direction
can be seen in Fig.~4. From Fig.~2b, it is
evident that a much stronger force is required for the unfolding 
when the force is applied along
the $y$-axis. The physical origin of this may be understood
from Fig.~5, where we have plotted schematic diagrams, keeping
$x=1$ (Fig.~5a) and $y=1$ (Fig.~5b), for a fixed extension of say $N/2$. 
It is easy to see that
in both cases, the number of contacts $\sigma$ is the same
($N/2$), while the
number of turns (or bends) in Fig.~5a is $2$ and in Fig.~5b, 
it is $N/2$.
As stiffness helps to stabilize the stretched state, the 
required force is less in the case of a force applied along the 
$x$-direction as compared to a force along the $y$-direction. 

In the constant force ensemble the control parameter $\alpha$
gets averaged, therefore, one does not find any oscillations 
(saw-tooth) in the control parameter in contrast to experiments \cite{rief,
busta,busta1,Oberhauser, Marszalek}. 
In a recent paper \cite{giri}, we have shown that the probability
distribution curve of the control parameter gives important information 
about the conformation of biomolecules in the form of 
oscillations in the control parameter. Keeping $f$ and $T$ 
constant near the transition line we plot, in Fig.~6, 
the probability distribution curves 
$P(| \alpha |)= (1/Z_N)\sum_{N_b,\sigma}
C_N (N_b, \sigma, | \alpha |) s^{N_b}
a^{\sigma} p^{| \alpha |} $ 
as a function of $\alpha$. 
Striking differences are apparent from these plots. When a 
force is applied either along the $x$- or $y$-axis, 
the probability distribution curve for flexible and
semi-flexible chains remain smooth at high temperatures.
However, at low temperature, when a force is applied
along $y$-axis, the emergence of peaks indicate the structural 
changes in biomolecules.  These features become more
apparent for semi-flexible chains, where we find peaks
at much higher temperature.  This may
be understood in the following way: If force is applied along
the $x$-axis, the loss of one monomer contact gives a unit of
extension along the $x$-axis. However, when force is
applied along the $y$-axis, there a loss of either one or 
two contacts always gives two units of extension along the
$y$-axis. This clearly 
shows that by changing the pulling direction one can obtain better
semi-microscopic information about  
the conformation of biomolecules. The features 
observed here should 
not be viewed as artifacts of the lattice model, because 
they appear only when the direction of pulling force is changed.

\begin{figure}[t]
\includegraphics[width=1.8in]{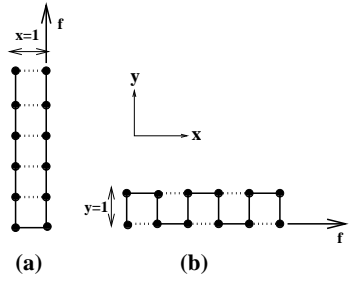}
\caption{\label{fig:model1} Schematic of two cases (having $N/2$ 
contacts and $N/2$ extension) where a force is applied along 
(a) the $y$-direction; (b) the $x$-direction. 
}
\end{figure}
  
\begin{figure}[t]
\includegraphics[width=3.in,height=4.in]{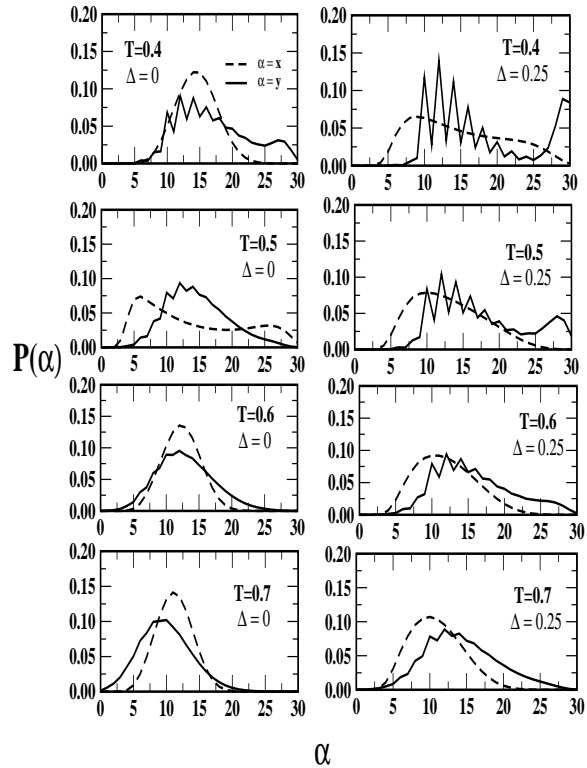}
\caption{\label{fig:prob} Probability distribution curve (dashed and 
solid line correspond to force along x and y respectively) at
different temperature for flexible chain ($\Delta=0$)
and semi-flexible chain ($\Delta=0.25$). 
}
\end{figure}

It is interesting to compare the qualitative features of our 
results with the ones obtained in experiments \cite{Dietz,brock}.
First, we note that  in one case (Refs~\cite{Dietz,brock}), the
system undergoes a shearing kind of transition for  which
the applied force is higher. This is the case (Fig.~1a) when a
force is applied along the $y$-axis. In the other case, 
it undergoes a $\beta$-sheet unzipping kind of transition
for which the applied force is less. This is reflected 
in Fig.~1(b), where we find that the critical force is less.
Our results provide strong evidence that saw-tooth like oscillations
are enhanced 
in force-extension curves due to the shearing (slippage) kind of 
transition. This can also be seen in experiments, where
the peaks of saw-tooth are much larger for shearing like transitions
compared to the unzipping of $\beta$-sheets. At this moment  
additional analytical and computer simulation work is 
required for a deeper understanding of the role of anisotropy
in the stability of biomolecules. 

It may be noted that the conformation of biomolecules remain 
unchanged by changing the direction of the force in the model discussed
here.  However, in experiments, by fixing one end only, the entire 
molecule will rotate due to the torque about the fixed end of the 
chain.  In order to see  effects predicted by 
the model, one has to fix not only the end point (as discussed 
above) but one more point in the chain. This will ensure that 
rotation will not take place around the fixed end due to the change 
in direction of the applied force.  A force may be applied at 
the other end of the chain either along the direction of the line 
connecting these points or perpendicular to this line.  By changing 
the position of the second point, one may get enhanced insight into the 
molecular interactions.

In conclusion, we have clearly demonstrated that finite-size
effects are crucial in understanding the experimental phase diagram
and that there are many intermediate states at low temperature.
Moreover, by considering a simple model which takes into account 
the anisotropy
of biomolecules, we have shown for the first 
time that changing the pulling direction gives distinct 
force-temperature curves even in the thermodynamic limit.
When a force is applied along the preferred direction, we observe 
the unzipping or opening of $\beta$-sheets layer by layer. 
However, when force is applied perpendicular to the 
preferred direction, we see the effects of slippage (shearing). 
It is evident from our studies that the mechanical resistance 
of biomolecules {e.g.} proteins, 
is not dictated solely by the amino acid sequence or unfolding rate
constant but depends critically on the topology of the biomolecules
and on the direction of the applied force.

We thank Haijun Zhou, Iwan Jensen and R. Rajesh for fruitful discussions 
on the subject and UGC, India for financial support.

\end{document}